\documentclass[11pt]{article}
\usepackage[totalheight = 23cm, totalwidth = 17cm]{geometry}
\usepackage{amssymb,amsmath,amsfonts,amsbsy,graphicx}
\usepackage{bm}

\def\mathbi#1{\textbf{\em #1}}

\newcommand{\mpl}{m_{\rm Pl}}
\newcommand{\fnl}{f_{\rm NL}}
\newcommand{\calR}{{\cal R}}

\begin{document}

\begin{titlepage}

\rightline{\footnotesize{APCTP-Pre2016-014}} 
\vspace{-0.2cm}
\rightline{\footnotesize{YITP-16-70}} 
\vspace{-0.2cm}

\begin{center}

\vskip 1.0 cm

{\LARGE \bf Consistency relation and inflaton field redefinition \\
 in the $\delta{N}$ formalism}

\vskip 1.0cm

{\large
Guillem Dom\`enech$^a$, Jinn-Ouk Gong$^{b,c}$ and Misao Sasaki$^{a}$
}

\vskip 0.5cm

\small{\it
$^{a}$Center for Gravitational Physics, Yukawa Institute for Theoretical Physics, 
\\
Kyoto University, Kyoto 606-8502, Japan
\\
$^{b}$Asia Pacific Center for Theoretical Physics, Pohang 37673, Korea
\\
$^{c}$Department of Physics, Postech, Pohang 37673, Korea
}

\vskip 1.2cm

\end{center}

\begin{abstract}

We compute for general single-field inflation the intrinsic non-Gaussianity
 due to the self-interactions of the inflaton field in the squeezed limit. 
We recover the consistency relation in the context of the $\delta{N}$ formalism, 
and argue that there is a particular field redefinition that makes the intrinsic 
non-Gaussianity vanishing, thus improving the estimate of the local 
non-Gaussianity using the $\delta{N}$ formalism.

\end{abstract}

\end{titlepage}

\newpage
\setcounter{page}{1}

\section{Introduction}

Now the paradigm of inflation is well established, its predictions have been
 tested against and survived the observations of the cosmic microwave background
 including the latest Planck mission~\cite{Ade:2015ava,Ade:2015lrj}. 
Ambitiously speaking, beyond being satisfied with the general idea of inflation, 
one now would like to be able to nail down the inflation model relevant for
 our observable universe by high-precision observations. 
Although the final answer might
 be still far in the future, current and planned observations are precise 
enough to start testing non-linear effects during inflation, beyond 
the power spectrum and spectral index~\cite{Ade:2015lrj}. In particular, 
non-Gaussianity has attracted a lot of attention, as it would easily 
constrain viable models of inflation~\cite{Ade:2015ava}.

To cope with high-precision observations, our theoretical understanding of
non-Gaussianity is required to be robust. It is thus important to bridge the 
remaining theoretical gaps and to check the consistency of the theory to 
estimate non-Gaussianity as accurate as possible. There are many works in
 this direction using the in-in formalism that allows us to compute $n$-point 
correlation functions of the primordial curvature perturbation. 
For extensive reviews of the in-in formalism and non-Gaussianity
 see, e.g.~\cite{nGreview,Koyama:2010xj} and references therein.

On the other hand, there is another powerful approach, the $\delta{N}$ 
formalism~\cite{deltaN}, to compute the $n$-point functions on super-horizon
 scales. The beauty of this formalism resides in its simplicity. Essentially,
one only needs to know the background evolution and the two-point function 
of the inflaton, given that the inflaton is {\em Gaussian}. 
However, a recent study by two of us~\cite{Gong:2015ypa} on k-inflation 
type general $P(X,\phi)$ theory with 
$X \equiv -g^{\mu\nu}\partial_\mu\phi\partial_\nu\phi/2$~\cite{k-inflation}
 pointed out that the $\delta{N}$ formalism may give large local non-Gaussianity,
 even though one assumes the usual slow-roll conditions and hence expects 
slow-roll suppression following the consistency relation of non-Gaussianity 
in the squeezed limit~\cite{nGconsistency,Li:2008gg}. 
However, it should be noted that such a result is based on the usual 
assumption of Gaussian inflaton field so that the intrinsic non-Gaussianity
 is not included. We can thus readily realize that we may find a different
 result if we think of, for example, a non-linear field redefinition.
 In other words, the notion of (non-)Gaussianity of the inflaton depends 
on how we define it, while the total non-Gaussianity does not.

The purpose of this article is twofold. First, we compute the intrinsic 
non-Gaussianity of the inflaton in the squeezed limit. Together with the 
naive estimate from the $\delta{N}$ formalism, we recover the consistency 
relation. Second, we argue that in attractor single field inflation there 
is a particular definition of the inflaton in which the intrinsic 
non-Gaussianity vanishes. This article is outlined as follows.
 In Section~\ref{sec:nG-deltaN}, we show that non-Gaussianity could be
 large in the $\delta{N}$ formalism, which however changes under field 
redefinition. Then, in Section~\ref{sec:int-nG}, we compute the three-point 
function of the inflaton fluctuation in the flat gauge and recover the
 consistency relation in the squeezed limit. Furthermore, 
in Section~\ref{sec:redefinition} we find a general field redefinition 
that removes the intrinsic non-Gaussianity for attractor.
 We summarize our results and conclude in Section~\ref{sec:conc}.

\section{Non-Gaussianity in the $\delta{N}$ formalism}
\label{sec:nG-deltaN}

Let us briefly recall the issue raised in~\cite{Gong:2015ypa}. 
Essentially, if one starts with k-inflation type theory and estimate 
non-Gaussianity with the $\delta{N}$ formalism, one finds an unexpected 
result with a new parameter. The action of our interest is
\begin{equation}
S = \int d^4x \sqrt{-g} \left[ \frac{\mpl^2}{2}R + P(X,\phi) \right] \, .
\end{equation}
This action yields the following background equations:
\begin{align}
\label{eq:BG-H}
& 3\mpl^2H^2 = 2XP_X - P \, ,
\\
\label{eq:dotH}
& \dot{H} = -\frac{XP_X}{\mpl^2} \, ,
\\
\label{eq:BGeom}
& \ddot\phi + 3H \left( 1 + \frac{p}{3} \right) \dot\phi = \frac{P_\phi}{P_X} \, ,
\end{align}
where we have defined
\begin{equation}
p \equiv \frac{\dot{P}_X}{HP_X} \, .
\end{equation}
Further, the speed of sound $c_s$ is given by $c_s^{-2} \equiv 1 + 2XP_{XX}/P_X$. 
Then, assuming that $H$ and $c_s$ are slowly varying, the spectral index of 
the power spectrum for the curvature perturbation $\calR$ is~\cite{k-inflation}
\begin{equation}
\label{eq:index}
n_\calR - 1 = -2\epsilon - \eta - s\, ,
\end{equation}
where 
\begin{align}
\epsilon \equiv -\frac{\dot{H}}{H^2} \, ,
\quad
\eta  \equiv \frac{\dot\epsilon}{H\epsilon} \, ,
\quad
s \equiv \frac{\dot{c}_s}{Hc_s} \, .
\end{align}
The observational constraint $n_\calR - 1 = 0.9645 \pm 0.0049$~\cite{Ade:2015lrj} 
tells us that these parameters should be much smaller than unity. 
Furthermore, using \eqref{eq:dotH}, we find an identity among these parameters,
\begin{equation}
\label{eq:eta}
\eta = 2(\epsilon+\delta) + p \, ,
\end{equation}
where we have defined
\begin{equation}
\delta \equiv \frac{\ddot\phi}{H\dot\phi} \, .
\end{equation}
Note that the spectral index \eqref{eq:index} only depends on $\epsilon$, $\eta$ and $s$.

Meanwhile, non-Gaussianity estimated by using the $\delta{N}$ formalism does 
depend on the other two parameters $\delta$ and $p$ as follows. 
During an attractor phase $N=N(\phi)$ we can write
\begin{equation}
\label{eq:deltaN}
\calR = \delta{N} = N_\phi\delta\phi + \frac{1}{2}N_{\phi\phi}\delta\phi^2 + \cdots \, ,
\end{equation}
where $\calR$ is the final comoving curvature perturbation and $\delta\phi$ 
is the inflaton fluctuation on the initial flat slice. Using
\begin{equation}
N_\phi = -\frac{H}{\dot\phi} \, ,
\end{equation}
we can rewrite \eqref{eq:deltaN} as
\begin{equation}
\label{eq:deltaN2}
\calR = -\frac{H}{\dot\phi}\delta\phi + \frac{1}{2}(\epsilon + \delta)
 \left( -\frac{H}{\dot\phi}\delta\phi \right)^2 + \cdots \, .
\end{equation}
Thus, identifying the linear, Gaussian component 
$\calR_g \equiv -H\delta\phi/\dot\phi$,
 the non-linear parameter $\fnl$ defined by~\cite{Komatsu:2001rj}
\begin{equation}
\label{eq:fNLdef}
\calR = \calR_g + \frac{3}{5}\fnl\calR_g^2
\end{equation}
can be read from \eqref{eq:deltaN2} as~\cite{fNL-deltaN}
\begin{equation}
\label{eq:fNLnaive}
\fnl^{\delta{N}} = \frac{5}{6}\frac{N_{\phi\phi}}{N_\phi^2}
 = \frac{5}{6}(\epsilon + \delta) = \frac{5}{12}(\eta - p) \, .
\end{equation}
It should be noted that nowhere through the derivation the smallness
of $p$ or $\delta$ is required. Thus, it would seem that we could have
large non-Gaussianity even in slow-roll inflation.

However, notice that $\epsilon$, $\eta$ and $c_s$ and in turn $s$ are 
invariant under a field redefinition, while $\delta$ and $p$ are not (see below).
This implies that our naive estimation \eqref{eq:fNLnaive} is not invariant 
under a field redefinition and, therefore, the intrinsic non-Gaussianity 
should account for extra information. Let us consider a general, non-linear 
field redefinition $\phi = f(\varphi)$. This means the fluctuations are related by
\begin{equation}
\delta\phi = f_\varphi\delta\varphi
 + \frac{1}{2}f_{\varphi\varphi}\delta\varphi^2 + \cdots \, ,
\end{equation}
so that even if (say) $\delta\varphi$ is Gaussian, $\delta\phi$ is not. 
Furthermore, $p(\phi)$ and $\delta(\phi)$ accordingly transform as, respectively,
\begin{align}
p(\phi) & = p(\varphi) - 2 \frac{\dot\varphi}{H}\frac{f_{\varphi\varphi}}{f_\varphi} \, ,
\\
\label{eq:del}
\delta(\phi) &
 = \delta(\varphi) + \frac{\dot\varphi}{H}\frac{f_{\varphi\varphi}}{f_\varphi} \, .
\end{align}
But $\delta{N}$ is invariant:
\begin{align}
\delta{N} & = \frac{1}{f_\varphi}N_\varphi\delta\phi + \frac{1}{2f_\varphi^2}
 \left( N_{\varphi\varphi} - \frac{f_{\varphi\varphi}}{f_\varphi}N_\varphi \right)
 \delta\phi^2 + \cdots
\nonumber\\
& = N_\varphi\delta\varphi + \frac{1}{2}N_{\varphi\varphi}\delta\varphi^2 + \cdots \,.
\end{align}
Despite being a trivial computation, we can derive an interesting implication.
The local non-Gaussianity \eqref{eq:fNLnaive} does change under such a 
transformation since in general $N_{\phi\phi}/N_\phi \neq N_{\varphi\varphi}/N_\varphi$.
As a result, the consistency relation in the squeezed limit seems to be violated.
We devote in the next section to show that the intrinsic non-Gaussianity of the inflaton 
completes the consistency relation and, in doing so, we look for an inflaton 
definition that minimizes the intrinsic non-Gaussianity.

\section{Intrinsic non-Gaussianity in the $\delta{N}$ formalism}
\label{sec:int-nG}

Let us take a rigorous look at the intrinsic non-Gaussianity in the context 
of the $\delta{N}$ formalism, in which we consider the local form of non-Gaussianity.
That is, we focus on the squeezed limit where one of the three modes has a 
wavelength much larger than the other two, say, $k_3 \ll k_1 \approx k_2$~\cite{fNL-deltaN}. 
In that limit, using \eqref{eq:deltaN} and \eqref{eq:fNLdef} give respectively 
the three-point function of the curvature perturbation as
\begin{align}
\left\langle \calR(\mathbi{k}_1)\calR(\mathbi{k}_2)\calR(\mathbi{k}_3) 
\right\rangle_{k_3 \ll k_1 \approx k_2} & 
= (2\pi)^3 \delta^{(3)}(\mathbi{k}_1 + \mathbi{k}_2 + \mathbi{k}_3) 
\left[ N_\phi^3 \left\langle \left\langle \delta\phi(\mathbi{k}_1)\delta\phi(\mathbi{k}_2) \right\rangle_{k_3} \delta\phi(\mathbi{k}_3) \right\rangle
+ 2\frac{N_{\phi\phi}}{N_\phi^2} P_\calR(k_1)P_\calR(k_3) \right]
\nonumber\\
& = (2\pi)^3 \delta^{(3)}(\mathbi{k}_1 + \mathbi{k}_2 + \mathbi{k}_3) 
\frac{12}{5} \fnl P_\calR(k_1)P_\calR(k_3) \, ,
\end{align}
where the subscript $k_3$ for the two-point function of $\delta\phi$ means that it is evaluated under the influence of 
the $k_3$ mode. An easy comparison tells us that the total non-linear parameter 
in the $\delta{N}$ formalism is given by
\begin{equation}
\fnl = \frac{5}{12} \left[ 2\frac{N_{\phi\phi}}{N^2_\phi} 
+ \frac{1}{N_\phi} 
\frac{\left\langle \left\langle \delta\phi(\mathbi{k}_1)\delta\phi(\mathbi{k}_2)
 \right\rangle_{k_3} \delta\phi(\mathbi{k}_3) \right\rangle}
{P_{\delta\phi}(k_1)P_{\delta\phi}(k_3)} \right] 
\equiv \fnl^{\delta{N}} + \fnl^\text{int} \, ,
\end{equation}
where $\fnl^{\delta{N}}$ is what we have found in \eqref{eq:fNLnaive}, 
$\fnl^\text{int}$ is due to the self-interaction of the inflaton, and 
we have used $P_\calR = N_\phi^2P_{\delta\phi}$. Thus we are left to compute
 the three-point function of the inflaton fluctuation in the squeezed limit, 
$\left\langle \left\langle \delta\phi(\mathbi{k}_1)\delta\phi(\mathbi{k}_2) 
\right\rangle_{k_3} \delta\phi(\mathbi{k}_3) \right\rangle$.

To compute $\fnl^\text{int}$, we proceed as follows. First, we keep
 the terms lowest order in slow-roll. One may worry that since we are 
interested in the case where $p$ and/or $\delta$ are not necessarily small, 
this may not be a good approximation. Nevertheless, we show later that
 we can actually redefine the inflaton and make $p$ and $\delta$ small. 
Second, following~\cite{long-short} we split $\delta\phi$ into long and 
short wavelength parts,
\begin{equation}
\delta\phi = \delta\phi_L + \delta\phi_S \, .
\end{equation}
In the squeezed limit, one of the modes has left the horizon long before 
the other two and therefore acts as background. That means we can keep the 
lowest order in $\delta\phi_L$ and neglect its spatial derivatives,
 since it is far outside the horizon. However, we do not neglect the time
 derivative of $\delta\phi_L$ because while $\calR$ is constant on 
super-horizon scales, $\delta\phi_L$ evolves as, to leading order,
\begin{equation}
\dot{\delta\phi}_L \approx -H(\epsilon + \delta) \frac{\dot\phi}{H}\calR
 = H(\epsilon+\delta)\delta\phi_L \, .
\end{equation}
It is thus essential to keep the time evolution of $\delta\phi_L$.

Without entering into details, the leading-order cubic action in the 
flat gauge~\cite{Koyama:2010xj} in the squeezed limit, after some algebra,
 is given by
\begin{align}
S_3 = \int d\tau d^3x & \Biggl[ \frac{a}{2} \frac{P_X}{\dot\phi}
 \delta\phi'_L \left\{ {\delta\phi_S'}^2 \left( {3} \left( c_s^{-2}-1 \right)
 + \frac{4X^2P_{XXX}}{P_X} \right)
 - (\nabla\delta\phi_S)^2 \left( c_s^{-2}-1 \right) \right\}
\nonumber\\
& + \frac{a^2}{2} \frac{HP_X}{\dot\phi} \delta\phi_L 
\Biggl\{ {\delta\phi_S'}^2 \left( c_s^{-2}p - 3\delta \left( c_s^{-2}-1 \right) 
- \epsilon\left(4c_s^{-2}-3\right) - 2c_s^{-2}s 
- 4(\epsilon+\delta)\frac{X^2P_{XXX}}{P_X} \right)
\nonumber\\
& \hspace{7em} + (\nabla\delta\phi_S)^2 
\left( \epsilon \left( c_s^{-2}-2 \right)
 -p + \delta \left( c_s^{-2}-1 \right) \right) \Biggr\} \Biggr] \, ,
\end{align}
where $d\tau = dt/a$ is the conformal time. Using this action, 
the in-in formalism yields
\begin{equation}
\frac{1}{N_\phi} 
\left\langle \left\langle \delta\phi(\mathbi{k}_1)\delta\phi(\mathbi{k}_2)
 \right\rangle_{k_3} \delta\phi(\mathbi{k}_3) \right\rangle
 = (2\epsilon+s+p) P_{\delta\phi}(k_1)P_{\delta\phi}(k_3) \, ,
\end{equation}
so that the intrinsic non-Gaussianity is
\begin{equation}
\label{eq:fNLint}
\fnl^\text{int} = \frac{5}{12} \left( 2\epsilon + s + p \right)
=\frac{5}{12} \left( \eta + s -2\delta\right)\, .
\end{equation}
Thus adding the above and $\fnl^{\delta{N}}$ given by (\ref{eq:fNLnaive})
together, we recover the consistency relation,
\begin{equation}
\label{eq:consistency}
\fnl =\frac{5}{12}(2\epsilon+\eta+s)
= \frac{5}{12} (1-n_\calR) \, .
\end{equation}

Importantly, it should be noted that the intrinsic 
non-Gaussianity \eqref{eq:fNLint} includes $p$ (or equivalently $\delta$), 
which is not invariant under a field redefinition. 
Therefore, if one could find a definition of the inflaton 
which minimizes \eqref{eq:fNLint}, that would be the perfect 
Gaussian definition for the inflaton. Let us show in the next section 
that this is possible in the attractor phase. Also we mention 
that one could use similar arguments presented in~\cite{Li:2008gg} in order 
to recover the consistency relation, although here our discussions are 
in the context of the $\delta{N}$ formalism.

\section{Most Gaussian definition of the inflaton}
\label{sec:redefinition}

Having found that $\fnl^{\delta{N}}$ and $\fnl^\text{int}$ include
the parameter $p$ which is dependent on the field redefinition while
$\fnl = \fnl^{\delta{N}} + \fnl^\text{int}$ is invariant, now in this section
we look for the field redefinition that leads to $\fnl^\text{int} = 0$ 
so that the $\delta{N}$ formalism alone gives improved estimate for the local 
non-Gaussianity. Before proceeding we first recall that \eqref{eq:eta} gives
\begin{equation}
p + 2\delta = \eta - 2\epsilon \ll 1 \,.
\end{equation}
Thus, minimizing $\delta$ minimizes $p$ at leading order approximation.

Now we start with a Lagrangian for $\varphi$ and work out the field 
redefinition $\phi = f(\varphi)$ that makes $\fnl^\text{int} = 0$. 
This is achieved if $2\delta = \eta + s$. 
Now, assuming an attractor phase, we can express the time derivative 
of $\varphi$ in terms of $\varphi$ as, say,
\begin{equation}
\label{eq:varphi-attractor}
\dot\varphi = g(\varphi) \, .
\end{equation}
Then \eqref{eq:del} is written as
\begin{equation}
\delta(\phi)
 = \frac{g}{H} \left( \frac{g_\varphi}{g}
 + \frac{f_{\varphi\varphi}}{f_\varphi} \right) \, .
\end{equation}
This implies that we can always choose the field redefinition 
$\phi=f(\varphi)$ that gives $\fnl^\text{int}=0$ by
\begin{equation}
\label{eq:f-deriv}
\log{f_\varphi} = -\log{g} + \beta(\varphi) + C \, ,
\end{equation}
where $C$ is an integration constant, and $\beta(\varphi)$ 
is determined by
\begin{equation}
\label{eq:beta-deriv}
\beta_\varphi = \frac{H}{g} \frac{\eta+s}{2} \, .
\end{equation}
Thus, given $g(\varphi)$, $H(\varphi)$, $\eta(\varphi)$ and $s(\varphi)$ 
we are able to find a solution for $\beta(\varphi)$. 
This in turn verifies that $p$ and $\delta$
can be always made small in attractor single field inflation
to render the system slow-rolling.

Let us consider a very simple example for illustration. 
Consider the following Lagrangian with $\alpha \gg 1$ and $\lambda \ll 1$:
\begin{equation}
\label{eq:pow}
P(Y,\varphi)
 = \frac{\lambda^2}{\alpha^2\varphi^2}Y - V_\star\varphi^{-\lambda^2/\alpha} \, ,
\end{equation}
where $Y \equiv -g^{\mu\nu}\partial_\mu\varphi\partial_\nu\varphi/2$.
 We can find that \eqref{eq:BG-H}, \eqref{eq:dotH} and \eqref{eq:BGeom} 
have an exact solution:
\begin{equation}
\varphi = \left(\frac{t}{t_0}\right)^{2\alpha/\lambda^2} \quad \text{and} 
\quad H = \frac{2}{\lambda^2t} \, ,
\end{equation}
where $t_0$ satisfies $V_\star t_0^2\lambda^4+2\lambda^2=12$. Therefore, we find
\begin{equation}
\epsilon = \frac{\lambda^2}{2} \, , \quad \eta = s = 0 \, , \quad \delta = \alpha - \frac{\lambda^2}{2} \quad \text{and} \quad p = -2\alpha \, .
\end{equation}
Note that $2\delta + p = -\lambda^2$, so that as long as $\lambda$ is small 
there is no  inconsistency with $\epsilon \ll 1$. 
However, $\varphi$ is fast rolling and the intrinsic non-Gaussianity 
could be as large as $\alpha$. 

By applying \eqref{eq:f-deriv}
 and \eqref{eq:beta-deriv} we can make $\fnl^\text{int} = 0$ as follows.
 First, we have
\begin{equation}
\dot\varphi = g(\varphi) = \frac{2\alpha}{\lambda^2t_0}\varphi^{1-\lambda^2/(2\alpha)} \, .
\end{equation}
Next we note that we may put $\beta=0$ from \eqref{eq:beta-deriv}.
Thus, \eqref{eq:f-deriv} gives
\begin{align}
f_\varphi=\frac{C}{g}=\frac{C\lambda^2t_0}{2\alpha}\varphi^{-1+\lambda/(2\alpha)}\,,
\end{align}
which, upon choosing $C=1/t_0$, may be integrated easily to give
\begin{equation}
\phi = f(\varphi) = \varphi^{\lambda^2/(2\alpha)} \,,
\end{equation}
apart from an irrelevant constant factor.
Under this redefinition the Lagrangian becomes
\begin{equation}
\label{eq:most-gaussian}
P(K,\phi) = \frac{4}{\lambda^2}K - \frac{V_\star}{\phi^{2}} \,;
\quad K\equiv-\frac{1}{2}g^{\mu\nu}\partial_\mu\log\phi\partial_\nu\log\phi\,.
\end{equation}
With this new field definition we see $\delta = 0$ so that 
$\fnl^\text{int} = 0$ as required, while $p = -\lambda^2$. 
Notice that such a field redefinition does not give a canonical
 kinetic term, contrary to what one would naively expect. 
The kinetic term in \eqref{eq:pow} becomes canonical by redefining
 the inflaton as
\begin{equation}
\psi = \frac{2}{\lambda}\log\phi=\frac{\lambda}{\alpha}\log\varphi \, ,
\end{equation}
leading to power-law inflation~\cite{Lucchin:1984yf}:
\begin{equation}
P(X,\psi) = X - V_\star e^{-\lambda\psi} \,;
\quad X=-\frac{1}{2}g^{\mu\nu}\partial_\mu\psi\partial_\nu\psi\,.
\end{equation}

It should be noted that the above procedure holds only when we have
{\em attractor} single field inflation and both $\epsilon$ and $\eta$ 
are small. 
If we do not demand $\eta \ll 1$ or attractor, the above procedure fails
to hold. 
For example, in the ultra-slow-roll inflation~\cite{usr} where $\eta = -6$, 
one cannot make $p$ and $\delta$ small at the same time by a field redefinition
 and the consistency relation is violated~\cite{non-attractor}.

\section{Discussions and conclusions}
\label{sec:conc}

There are certain cases in slow-roll attractor single field inflation that 
the $\delta{N}$ formalism gives a large local non-Gaussianity \eqref{eq:fNLnaive}. 
This contradicts the consistency relation that it should be slow-roll suppressed. 
The reason for this inconsistency is that the $\delta{N}$ formalism assumes that 
the inflaton is Gaussian. We have explicitly checked that by properly taking 
into account the intrinsic non-Gaussianity \eqref{eq:fNLint} the consistency 
relation is recovered as \eqref{eq:consistency}. Moreover, since the notion of 
Gaussianity is sensitive to a non-linear field redefinition while the total 
non-Gaussianity is not, we have found the field 
redefinition \eqref{eq:varphi-attractor} and \eqref{eq:f-deriv} that makes 
the intrinsic non-Gaussianity \eqref{eq:fNLint} vanish in the squeezed limit. 
Interestingly, the most Gaussian field definition needs not coincide with
 a canonical field as can be seen in \eqref{eq:most-gaussian}.

Throughout this article, we have worked under the assumption that we are 
in attractor phase. However, in non-attractor inflationary models where the 
consistency relation does not hold there is no reason to anticipate that 
either (or both) the naive estimate of non-Gaussianity from the $\delta{N}$ 
formalism or (and) the intrinsic non-Gaussianity should be small. 
Although it is out of the scope of the present article, it would be interesting 
to study the role of $p$ and $\delta$ in non-attractor models. 
Finally, it is worthwhile to mention that since the consistency relation is 
general, one could easily generalize the most Gaussian inflation definition 
to more general scalar-tensor theories of gravity.

\subsection*{Acknowledgements}

We would like to thank Joey Fedrow, Antonio de Felice, Godfrey Leung, 
Atsushi Naruko, Ryo Saito, Yuki Sakakihara and Takahiro Tanaka for useful 
comments and discussions. 
JG acknowledges support from the Korea Ministry of Education, Science and 
Technology, Gyeongsangbuk-Do and Pohang City for Independent Junior Research 
Groups at the Asia Pacific Center for Theoretical Physics.
This work is also supported in part by MEXT KAKENHI Grant Number 15H05888, 
by a Starting Grant through the Basic Science Research Program of the National 
Research Foundation of Korea (2013R1A1A1006701) and by a TJ Park Science 
Fellowship of POSCO TJ Park Foundation.

\newpage

\appendix

\section{Simplifying the third order action}
\label{app:S3}

Here, we present a series of simplifications and formulae that we use in the main text. The third order action in the flat gauge is given by~\cite{Koyama:2010xj}
\begin{align}
\label{eq:third}
S_3 = \int d^4x a^3 & \left[ P_{XX} \left\{ \frac{1}{2}\dot\phi\dot{\delta\phi}^3 + X\alpha \left( -4\dot{\delta\phi}^2 + 5\dot\phi\alpha\dot{\delta\phi} - 4X\alpha^2 \right) \right. \right.
\nonumber\\
& \hspace{3em} \left. + a^{-2} \left[ -\frac{1}{2}\dot\phi\dot{\delta\phi}(\nabla\delta\phi)^2 + X\alpha(\nabla\delta\phi)^2 - 2X \left( \dot{\delta\phi} - \dot\phi\alpha \right)\partial_i\delta\phi\partial^i\psi \right] \right\}
\nonumber\\
& + P_{X\phi} \left[ \frac{1}{2}\delta\phi\dot{\delta\phi}^2 - \dot\phi\alpha\delta\phi\dot{\delta\phi} + X\alpha^2\delta\phi - a^{-2} \left(\frac{1}{2}\delta\phi(
\nabla\delta\phi)^2 + \dot\phi\delta\phi\partial_i\delta\phi\partial^i\psi\right) \right]
\nonumber\\
& + P_{XXX} X \left[ \frac{1}{3}\dot\phi\dot{\delta\phi}^3 + X\alpha \left( -2\dot{\delta\phi}^2 + 2\dot\phi\alpha\dot{\delta\phi} - \frac{4}{3}X\alpha^2 \right) \right]
\nonumber\\
& + P_{XX\phi}X \left( \delta\phi\dot{\delta\phi}^2 - 2\dot\phi\alpha\delta\phi\dot{\delta\phi} + 2X\alpha^2\delta\phi \right) + P_{X\phi\phi} \left( \frac{1}{2}\dot\phi\dot{\delta\phi} - X\alpha \right) \delta\phi^2
\nonumber\\
& + P_X \left\{ \alpha \left( -\frac{1}{2}\dot{\delta\phi}^2 + \dot\phi\alpha\dot{\delta\phi} - X\alpha^2 \right) - a^{-2} \left[ \frac{1}{2}\alpha(
\nabla\delta\phi)^2 + \left( \dot{\delta\phi} - \dot\phi\alpha \right) \partial_i\delta\phi\partial^i\psi \right] \right\}
\nonumber\\
& \left. + \frac{1}{2}P_{\phi\phi}\alpha\delta\phi^2 + \frac{1}{6}P_{\phi\phi\phi}\delta\phi^3 + 3H^2\alpha^3 + H\alpha^2\frac{\Delta}{a^2}\psi + \frac{1}{2a^4}\alpha \left[ (\Delta\psi)^2 - \partial_i\partial_j\psi\partial^i\partial^j\psi \right] \right] \, ,
\end{align}	
where $\alpha$ and $\psi$ are given respectively by
\begin{align}
\alpha & = \epsilon H \frac{\delta\phi}{\dot\phi} \, ,
\\
\Delta\psi & = \frac{a^2\epsilon}{c_s^2} \frac{d}{dt} \left( -\frac{H}{\dot\phi}\delta\phi \right) \, .
\end{align}

Keeping the leading slow-roll terms simplifies the action to yield
\begin{align}
S_3 = \int d^4x \frac{a^3P_X}{\dot\phi} & \left\{ {\delta\dot\phi}^3 \left( \frac{XP_{XX}}{P_X} + \frac{2}{3}\frac{X^2P_{XXX}}{P_X} \right) \right.
\nonumber\\
& + H{\delta\dot\phi}^2\delta\phi \left[ \frac{\dot\phi P_{XX\phi}X}{HP_X} + \frac{1}{2}\frac{\dot\phi P_{X\phi}}{HP_X} - \frac{1}{2}\epsilon \left( 1 + \frac{8XP_{XX}}{P_X} + \frac{4X^2P_{XXX}}{P_X} \right) \right]
\nonumber\\
& + H^2{\delta\dot\phi}\delta\phi^2 \left[ \frac{XP_{X\phi\phi}}{H^2P_X} - \epsilon \left( \frac{1}{2}\frac{\dot\phi P_{X\phi}}{HP_X} \left( 2+c_s^{-2} \right) + \frac{2\dot\phi P_{X\phi}X}{HP_X} \right) \right]
\nonumber\\
& + H^3\delta\phi^3 \left[ \frac{1}{6}\frac{\dot\phi P_{\phi\phi\phi}}{H^3P_X} + \epsilon \left( \frac{1}{2}\frac{P_{\phi\phi}}{H^2P_X} - \frac{XP_{X\phi\phi}}{H^2P_X} \right) \right] + \frac{\epsilon}{c_s^2} {\delta\dot\phi}\partial^i\delta\phi\Delta^{-1}\partial_i\dot{\delta\phi} \left( 1+\frac{2XP_{XX}}{P_X} \right)
\nonumber\\
& \left. - a^{-2}\delta\dot\phi(\nabla\delta\phi)^2\frac{XP_{XX}}{P_X}+ \frac{1}{2a^2}H\delta\phi(\nabla\delta\phi)^2 \left[ \epsilon \left( \frac{2XP_{XX}}{P_X} - 1 \right) - \frac{P_{X\phi}\dot\phi}{HP_X} \right] \right\} \, .
\end{align}
If we further keep only leading order in all the parameters introduced in Section~\ref{sec:nG-deltaN}, we find
\begin{align}
\label{eq:3rd}
S_3 = \int d^4x \frac{1}{2}\frac{a^3P_X}{\dot\phi} & \left\{ \dot{\delta\phi}^3 \left[ \left( c_s^{-2}-1 \right) + \frac{4}{3}\frac{X^2P_{XXX}}{P_X} \right] \right.
\nonumber\\
& + H\dot{\delta\phi}^2\delta\phi \left[ c_s^{-2}p - 3\delta \left( c_s^{-2}-1 \right) - 2c_s^{-2}s - (\delta+\epsilon)\frac{4X^2P_{XXX}}{P_X} - \epsilon\left( 4c^{-2}_s-3 \right) \right]
\nonumber\\
& - a^{-2}\dot{\delta\phi}(\nabla\delta\phi)^2 \left( c_s^{-2}-1 \right) + 2\epsilon\dot{\delta\phi}\partial^i\delta\phi\Delta^{-1}\partial_i\dot{\delta\phi} 
\nonumber\\
& + a^{-2}H\delta\phi(\nabla\delta\phi)^2 \left[ \epsilon \left( c_s^{-2}-2 \right) - p + \delta \left( c_s^{-2}-1 \right) \right] \Bigg\} \, ,
\end{align}
where we have expressed the coefficients in the third order action in terms of the parameters defined in Section~\ref{sec:nG-deltaN} using the following  relations:
\begin{align}
\frac{\dot{\phi}P_{X\phi}}{HP_X} & = p + \delta \left( 1 - c_s^{-2} \right) \, ,
\\
\frac{{P_{\phi\phi}}}{H^2P_X} & = \frac{\dot{p}+\dot{\delta}}{H} + (p+\delta+3)(p+\delta-\epsilon) - \delta \left[ p + \delta \left( 1-c_s^{-2} \right) \right] \, ,
\\
\frac{2\dot\phi P_{XX\phi}X}{HP_X} & = \left( c_s^{-2}-1 \right)(p-2\delta) - 2c_s^{-2}s - \delta\frac{4X^2P_{XXX}}{P_X} \, ,
\\
\frac{2XP_{X\phi\phi}}{H^2P_X} & = \frac{\dot{p}+\dot{\delta} \left( 1-c_s^{-2} \right)}{H} + 2c_s^{-2}s\delta + (p-\epsilon-\delta) \left[ p + \delta \left( 1-c_s^{-2} \right) \right] - \delta \frac{2\dot{\phi}P_{XX\phi}X}{HP_X} \, ,
\\
\frac{{P_{\phi\phi\phi}\dot\phi}}{H^3P_X} & = (p-2\epsilon) \frac{{P_{\phi\phi}}}{H^2P_X} - \delta\frac{2XP_{X\phi\phi}}{H^2P_X} + \frac{\ddot{p}+\ddot{\delta}}{H^2} + (2p+\delta)\frac{\dot{p}}{H} + p\frac{\dot{\delta}}{H} - \epsilon\eta(p+\delta)
\nonumber\\
& \quad - \frac{\dot\delta}{H} \left[ p + \delta \left( 1-c_s^{-2} \right) \right] + 2c_s^{-2}\delta\left( \frac{\dot\delta}{H} + s \right) \, .
\end{align}

\end{document}